\documentclass[showpacs,twocolumn,prb,amssymb,amsmath,nobibnotes,aps]{revtex4}
\usepackage{bm}

\newcommand{\be}{\begin{equation}}    \newcommand{\ee}{\end{equation}}
\newcommand{\bea}{\begin{eqnarray}}
\newcommand{\eea}{\end{eqnarray}}
\newcommand{\half}{\frac{1}{2}}

\newcommand{\nh}{\hat{n}}
\newcommand{\zh}{\hat{z}}
\newcommand{\xv}{{\bf x}}
\newcommand{\rv}{{\bf r}}

\newcommand{\sn}{\mathrm{sn}}
\def\rf#1{(\ref{#1})}
\usepackage[dvips]{graphicx}

\begin{document}

\title{Conical soliton escape into a third dimension of a surface vortex}
\author{Leo Radzihovsky}
\author{Quan Zhang}
\affiliation{Department of Physics, University of Colorado,
   Boulder, Colorado 80309, USA}

\date{\today}

\begin{abstract}
  We present an exact three-dimensional solitonic solution to a sine-Gordon-type Euler-Lagrange equation, that describes a
  configuration of a three-dimensional vector field $\nh$ constrained to a surface $p$-vortex, with a prescribed polar tilt angle on a planar substrate and escaping into the third dimension in the bulk. The solution is relevant to characterization of a schlieren texture in nematic liquid-crystal films with tangential (in-plane) substrate alignment. The solution is identical to a section of a point defect discovered many years ago by Saupe [Mol. Cryst. Liq. Cryst. {\bf 21}, 211 (1973)], when latter is restricted to a surface.
\end{abstract}
\maketitle

\section{Introduction}
\label{introduction}

Topological defects are central to a complete description of
ordered phases of condensed matter, ranging from superconductors to liquid crystals\cite{deGennes}. Defects' energetics controls the stability of the ordered state to thermal fluctuations\cite{KT73,HNY78}, random material heterogeneities\cite{CO82} and external perturbations\cite{ChaikinLubensky}. A complete rigorous classification\cite{Mermin79} is now available for most {\em bulk} ordered states.  

This has been particularly fruitful in understanding a rich variety of topological defects that are found in liquid-crystal phases. However, in many physical contexts, as, for example, arising in liquid crystals confined inside a thin display cell, much of the physics is controlled by a substrate interaction which competes with the bulk energetics\cite{deGennes}. In such surface-dominated situations, only an incomplete understanding of defects structure and stability is available.

One important and extensively studied example of this type discovered by Meyer\cite{Meyer73} is that of a uniaxial nematic liquid crystal confined to a thin long capillary with a homeotropic alignment at the cylindrical surface. The resulting boundary condition forces an integer winding of the nematic director field, which for a two-dimensional ({\em xy}) field would trap a vortex line along the axis of the capillary. However, such defect is unstable for a three-dimensional (3D) director field and away from the boundary exhibits an escape into the third dimension, removing the line
singularity as described by Meyer's solution\cite{Meyer73}.

A familiar schlieren surface texture seen in phase contrast microscopy is a hallmark of nematic liquid crystals, reflecting surface-induced disclinations (vortices in the nematic director field). The texture details, e.g., appearance of integer versus half-integer vortices have been suggested to distinguish between the uniaxial and biaxial nematic states\cite{Chandrasekhar98,Chiccoli02}. New advanced bulk imaging techniques, such as, for example, the fluorescence confocal polarizing microscopy, have also allowed imaging of the full three-dimensional textures associated with such surface defects\cite{Smalyukh04}.
\begin{figure}[bth]
\centering
\setlength{\unitlength}{1mm}
\begin{picture}(0,80)(0,0)
\put(-52,-7){\begin{picture}(0,60)(0,0)
\includegraphics{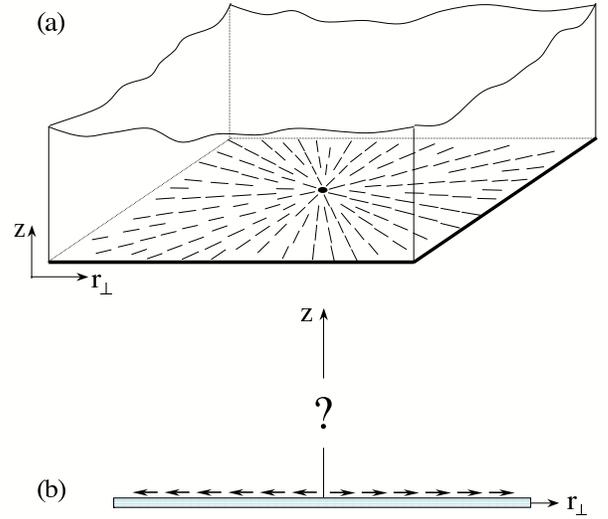}
\end{picture}}
\end{picture}
\caption{A nematic director field $\nh(\xv)$ constrained to a $2\pi$-vortex on a surface $z=0$ with a planar alignment rendered in 3D in (a) and in a 2D projection in (b). The corresponding bulk texture configuration $\nh(\xv)$ that minimizes the Frank energy is calculated analytically in this paper and is illustrated in Fig.\rf{SVsolution2DFig}.}
\label{SVproblemFig}
\end{figure}

Motivated by the above discussion, here we consider a problem of a 3D nematic with a planar (parallel) surface alignment, with an integer vortex imposed on a substrate, as illustrated in Fig.\rf{SVproblemFig}. In contrast to the long capillary
case\cite{Meyer73}, that clearly exhibits translational invariance along its axis, reducing it to one dimension (1D), here the system is manifestly three-dimensional [two-dimensional (2D), once azimuthal symmetry is included], and therefore in principle considerably more complicated.

Here we present a derivation of an {\em exact} solution to the single Frank elastic constant Euler-Lagrange (E-L) equation that describes a bulk texture induced by a surface $2\pi p$-vortex, with $p$ the integer azimuthal vortex winding number. It is described in terms of the polar angle $\theta(r_\perp,z)=\theta_s(z/r_\perp)$ of the director field $\nh(\xv)$ that we find to be given by
\begin{equation}
\theta_s(t)=
2\mathrm{arccot}\left[\left(t+\sqrt{t^2+1}\right)^p\right].
\label{solitonSolutionIntro}
\end{equation}
Illustrated in Fig.\rf{SVsolution2DFig} for $p=1$, the bulk texture is a conical soliton giving the nematic director's escape into the third dimension away from the imposed surface $2\pi p$-vortex with a strong planar alignment.  
\begin{figure}[bth]
\centering
\setlength{\unitlength}{1mm}
\begin{picture}(30,60)(0,0)
\put(-33,-8){\begin{picture}(30,60)(0,0)
\includegraphics{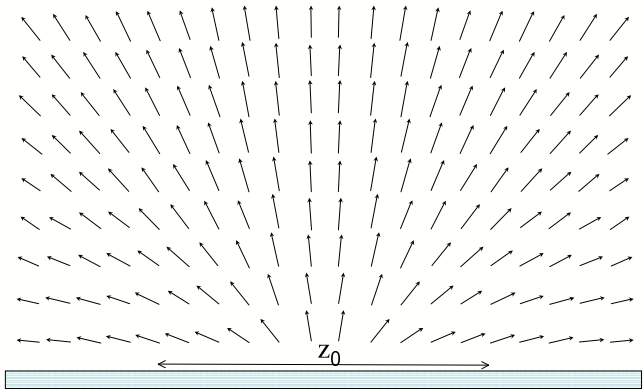}
\end{picture}}
\end{picture}
\caption{Exact conical soliton solution (here projected onto $z-r_\perp$ plane) of the nematic director field, $\nh(\xv)$, describing escape into the third dimension of a surface $2\pi$-vortex.}
\label{SVsolution2DFig}
\end{figure}

The director $\nh(\xv)$ configuration corresponding to Eq.\rf{solitonSolutionIntro} coincides with a section of the Saupe's point $p$-defect\cite{Saupe} [$\tan\theta/2=(\tan\delta/2)^{|p|}$, with $\delta$ as a polar angle of
the spherical coordinates] when restricted to a subspace above a planar substrate. For $p=1$ the above result reduces to a well-known simple texture, given by half of the hedgehog (skyrmion) configuration, $\nh(\xv)=\hat{\xv}$. Application of Saupe's $p=1$-defect to a surface vortex problem was also previously explored by Kleman\cite{Kleman} and was shown to satisfy the simplest homogeneous boundary conditions arising from a model pinning potential\cite{Kleman}.

To summarize our contributions, we present a derivation (in cylindrical coordinates, mapping the Euler-Lagrange equation to that of a dissipative particle with a time-dependent mass) of Saupe's solution [Eq.\rf{solitonSolutionIntro}]. This approach is likely extendable (even if approximately) to a study of other interesting problems of surface defects, where the reduction to Saupe's point defect solution no longer holds.  Our slight generalization of Saupe's solution allows us to discuss and connect to weak and strong anchorings. Finally, we present an analysis of the energetics, comparing the conical soliton escape to other competing textures. We discover a counterintuitive dependence of the $p$-vortex energy on $p$, showing that it is asymptotically linear in $p$, in contrast to the standard $p^2$ dependence. This has important implications for the stability of $p>1$ surface vortices over their fission into $p$ lower winding ($p=1$) vortices.

\section{Model}

We consider a 3D model of a nematic liquid crystal, characterized by a nematic unit director field, $\nh(\xv)$, with $\xv=(\rv_\perp,z)$. The energy is given by a Hamiltonian
\begin{equation}
H = H_{el} + H_{s},
\label{H}
\end{equation}
where $H_{el}$ is the bulk elastic energy of the Frank model
\begin{eqnarray}
\hspace{-0.4cm}
H_{el}&=&\frac{1}{2}\int d^2\rv_\perp dz\bigg[K_1(\nabla\cdot\nh)^2 +K_2[\nh\cdot(\nabla\times\nh)]^2\nonumber\\
 &&+ K_3[\nh\times(\nabla\times\nh)]^2\bigg],
\label{H_Frank}
\end{eqnarray}
and $H_s$ is the surface pinning energy, localized at $z=0$,
\begin{equation}
H_{s}=\int d^2\rv_\perp dz V_s(\rv_\perp) \delta(z)(\zh\cdot\nh)^2,
\label{Hs}
\end{equation}
with $\zh$ as the surface normal. In the single elastic constant approximation, $K_1=K_2=K_3=K$, the elastic energy reduces to
\begin{equation}
H_{el}=\frac{K}{2}\int d^2\rv_\perp dz (\nabla\nh)^2.
\label{H_singleK}
\end{equation}
In above we have dropped the boundary terms as they do not affect the E-L equation. Parametrizing the unit director field 
\begin{equation}
\nh=(\sin\theta\cos\phi,\sin\theta\sin\phi,\cos\theta)
\label{nhat}
\end{equation}
in terms of polar and azimuthal angles $\theta$ and $\phi$, $H_{el}$
reduces to
\begin{equation}
H_{el}=\frac{K}{2}\int d^2\rv_\perp dz \bigg[(\nabla\theta)^2 
+ \sin^2\theta(\nabla\phi)^2\bigg].
\label{H_thetaphi}
\end{equation}
We can include surface pinning through a boundary condition on
$\nh(\rv_\perp,z=0)=\nh_0(\rv_\perp)$, finding the corresponding solution and then minimizing over $\nh_0(\rv_\perp)$ in the presence of $V_s(\rv_\perp)$.

We focus on the solution $\nh(\xv)$ subject to a constraint of a $2\pi p$-vortex ($p\in{\cal Z}$) at $z=0$. The $2\pi p$ surface winding is imposed by taking 
\begin{equation}
\phi(\rv_\perp,z=0)=p\varphi+\varphi_0, 
\label{pvarphi}
\end{equation}
where $\varphi=\arctan(y/x)$ is the azimuthal angle of the cylindrical coordinate system $\xv=(r_\perp\cos\varphi,r_\perp\sin\varphi,z)$. The arbitrary constant angle $\varphi_0$ gives a family of textures induced by spiral surface defects for $0<\varphi_0<\pi/2$. These interpolate between a pure splay surface ``aster'' defect for $\varphi_0=0$, [illustrated in Fig.\rf{SVproblemFig}(a)] and a pure bend surface ``vortex'' defect for $\varphi_0=\pi/2$.

The resulting elastic energy is then given by
\begin{equation}
H_{el}=\frac{K}{2}\int d^2\rv_\perp dz \bigg[(\nabla\theta)^2 
+ \frac{p^2}{r_\perp^2}\sin^2\theta\bigg],
\label{H_vortex}
\end{equation}
leading to the Euler-Lagrange equation that determines the texture configuration $\theta(\xv)$,
\begin{equation}
\nabla^2\theta - \frac{p^2}{2 r_\perp^2}\sin2\theta = 0.
\label{EL}
\end{equation}

Focusing for simplicity on azimuthally symmetric boundary conditions, we search for a $\varphi$-independent solution $\theta(r,z)$,
satisfying
\begin{equation}
r^2\partial_r^2\theta + r\partial_r\theta + r^2\partial_z^2\theta-\frac{p^2}{2}\sin2\theta = 0,
\label{ELrz}
\end{equation}
with a surface constraint $\theta(r,z=0)=\theta_0(r)$. We have simplified the notation by denoting $r_\perp\equiv r$.

\section{Solution of the Euler-Lagrange equation}
\label{solutionEL}

Despite the fact that the E-L equation, Eq.\rf{ELrz} is nonlinear and two-dimensional, its one- (and periodic array-) soliton solution can be found exactly\cite{Saupe}. The intuition for the form of the solution can be obtained by neglecting the $r$-derivative terms and then noting that the resulting equation is of a standard 1D sine-Gordon type along $z$, with a period $\pi$. It thus admits a soliton solution connecting tilt angle $\theta(r,z=-\infty)=\pi$ to
$\theta(r,z=+\infty)=0$ with the soliton width at the transverse distance $r$ from the vortex given by $\xi_z(r) = r$.

\subsection{Exact conical soliton solution}

Motivated by the above observation and by the translational invariance of the E-L equation along $z$, we search for a soliton solution of the form
\begin{equation}
\theta(r,z)\equiv\theta_s\left(\frac{z+z_0}{r}\right).
\end{equation}
We note that this restricted form precludes a study of other than a constant boundary condition at $z=-z_0$. Since in general the symmetry dictates a nontrivial radial variation in the director tilt angle at the surface with the distance $r$ from the vortex, we anticipate that the above form of the solution is an exact description only for an infinitely strong planar alignment on a substrate at $z=-z_0=0$. For a
finite planar surface anchoring, given large azimuthal strain near the vortex, we expect a meron configuration with $\theta_0(r)\approx0$ in the vicinity of the vortex (near $\rv=0$) and growing to $\pi/2$ with increasing distance $r$ from it. As we will see below, the radial surface variation $\theta_0(r)\equiv\theta(r,z=0)$ can be qualitatively captured by the solution $\theta_s(z_0/r)$ at $z=0$, by adjusting $z_0$. While this single degree of freedom ($z_0$) is in principle insufficient to capture an arbitrary form of the surface boundary condition, $\theta_0(r)$, we proceed to explore this class of solutions\cite{commentKleman}. We expect it to be a good approximation for strong planar anchoring, characterized by a vanishing $z_0$ and corresponding to $\theta_0(r)\approx\pi/2$ for nearly all $r$,
excluding a small core region of radius $z_0$. We will treat the weakly anchored case in a complementary way.

As a function of the scaling variable $t=(z+z_0)/r$, the E-L equation simplifies to
\begin{equation}
m(t)\ddot{\theta}_s + \gamma(t)\dot{\theta}_s - \frac{p^2}{2}\sin2\theta_s=0,
\label{newton}
\end{equation}
where
\begin{eqnarray}
m(t)&=& t^2+1,\\
\gamma(t)&=& t.
\label{m_gamma}
\end{eqnarray}

Solutions of Eq.\rf{newton} can be most easily obtained by its
identification with Newton's equation for a particle at position $\theta(t)$ at time $t$, moving in a periodic potential $V(\theta)=\frac{p^2}{4}\cos2\theta$ and characterized by time-dependent mass and friction coefficients, $m(t)$ and $\gamma(t)$, respectively. This type of identification is quite analogous to a standard sine-Gordon model, where, in contrast, the fictitious particle has a constant mass and no friction. In this latter case the solution is easily obtained by a guaranteed existence of an integral of motion, energy of the particle, which reduces the solution to a single integral. In our problem the time dependence of the mass and finite friction at first sight would be expected to preclude the existence of such ``conservation of energy'' integral of motion. However, a key observation is that the two effects can exactly compensate each other if the condition
\begin{equation}
\gamma(t)=\half\dot{m}(t)
\end{equation}
is satisfied (as it is in our problem), and leads to an ``energy'' conservation law
\begin{equation}
\frac{d}{d t}\bigg[\half m(t)\dot{\theta_s}^2 + 
\frac{p^2}{4}\cos2\theta_s\bigg]=0.
\label{conserveE}
\end{equation}
Indeed this is, guaranteed by the fact that the E-L Eq.\rf{EL} came from a minimization (of $H_{el}$) principle.

The resulting integral of ``motion,'' 
\begin{equation}
\half m(t)\dot{\theta}_s^2 + \frac{p^2}{4}\cos2\theta_s=\frac{p^2}{4}E,
\label{conserveE2}
\end{equation}
with $p^2E/4$ as the fictitious particle's energy, then easily allows us to reduce the problem to a single integral
\begin{equation}
\int_{t_0}^tdt'\frac{p}{\sqrt{m(t')}} = 
-\sqrt{2}\int_{\pi/2}^{\theta_s}\frac{d\theta'}{\sqrt{E-\cos2\theta'}},
\label{integralSoln}
\end{equation}
where in above, we have made a choice of the negative square-root. The parameter $t_0$ defined by $\theta_s(t_0)=\pi/2$ determines the tilt angle $\theta_0$ at the boundary at $z=-z_0$. The other constant of ``motion,'' $E$, is also crucial to the nature of the solution. It is quite clear that for a half-infinite space boundary conditions (see Fig.\rf{SVproblemFig}) $E$ must be chosen to be $E=1^-$ so that the
solution is a single soliton in $t$. In the mechanics analogy it corresponds to a particle at $t=-\infty$, starting out at
$\theta=\pi$, with the energy just equal to the potential energy, i.e., with an infinitesimally vanishing initial velocity, rolling down the hill during $-\infty < t < t_0$ and then climbing back up to the top at $\theta=0$ as $t\rightarrow +\infty$.

Using $m(t)=t^2+1$, Eq.\rf{integralSoln} is easily integrated,
\begin{eqnarray}
\int_{t_0}^tdt'\frac{p}{\sqrt{t'^2+1}} &=& 
-\int_{\pi/2}^{\theta_s}\frac{d\theta'}{\sin\theta'},\nonumber\\
p\ln\left[\frac{t+\sqrt{t^2+1}}{t_0+\sqrt{t_0^2+1}}\right]&=& \ln\cot(\theta_s/2),
\label{integrate}
\end{eqnarray}
and leads to our main result
\begin{equation}
\theta_s(t)=
2\mathrm{arccot}
\left[\left(\frac{t+\sqrt{t^2+1}}{t_0+\sqrt{t_0^2+1}}\right)^p\right],
\label{solitonSolution}
\end{equation}
illustrated in Fig.\rf{solitonFig}. The corresponding director field texture is illustrated in Fig.\rf{SVsolution2DFig}.
\begin{figure}
\begin{picture}(0,0)(0,0)
\put(180,120){(a)}
\end{picture}
  \includegraphics[height=2 in]{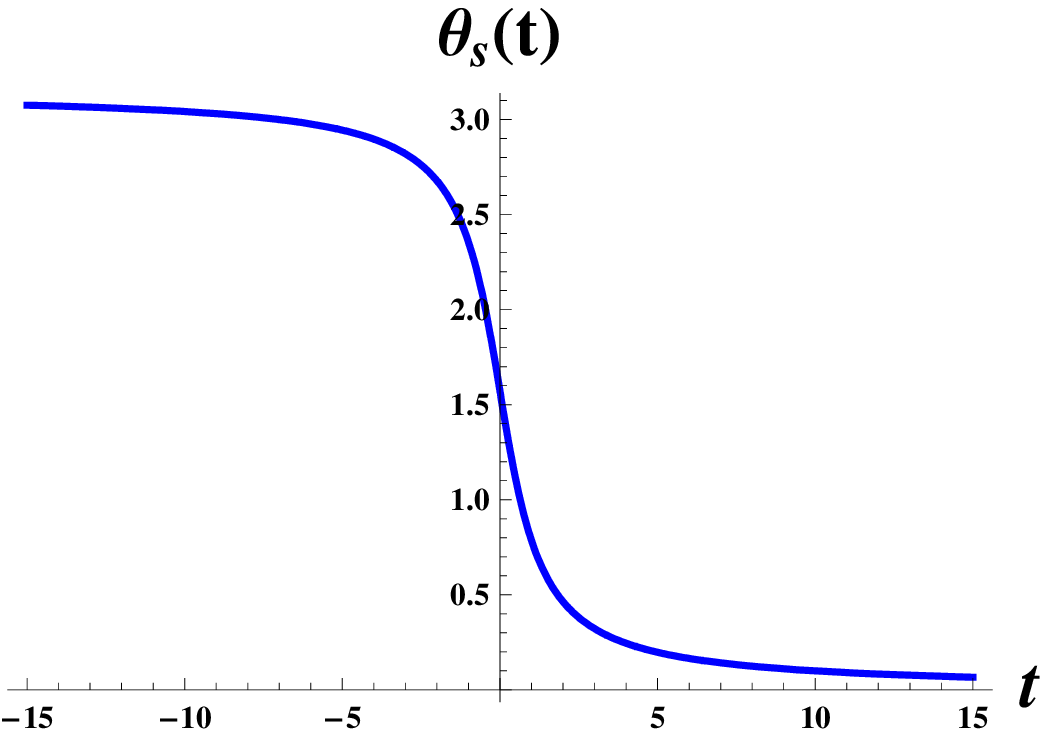}
\vspace{0.5cm}

\begin{picture}(0,0)(0,0)
\put(180,120){(b)}
\end{picture}
  \includegraphics[height=2 in]{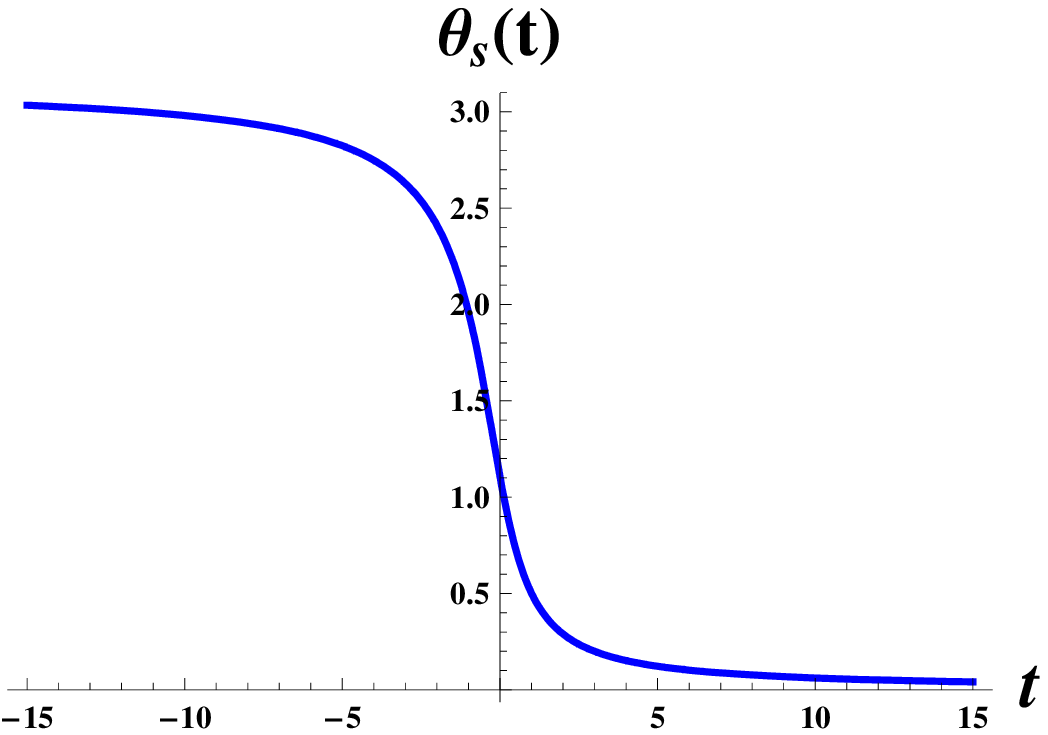}\\       
  \caption{Solitonic $p=1$ solution from Eq.\rf{solitonSolution} for (a) $t_0=0$, corresponding to a perfect planar alignment, $\theta_0=\pi/2$ at $z=-z_0$, and (b) $t_0=-0.5$, corresponding to a uniform tilt of $\theta_0=1.11$ radians at the $z=-z_0$ boundary.}
\label{solitonFig}
\end{figure}
The solution satisfies the Euler-Lagrange equation with a uniform tilt $\theta_0$ boundary condition,
\begin{equation}
\theta(r,z=-z_0)=\theta_0=2\arctan\left[\left(t_0+\sqrt{t_0^2+1}\right)^p\right]
\label{boundary_z0}
\end{equation} 
on the $z=-z_0$ surface. Clearly, $\theta_0$ is also the asymptotic tilt angle at large $r$ (vanishing $t$), and in terms of it, the solution can be equivalently written as
\begin{equation}
\theta_s(t)=
2\mathrm{arccot}\left[\cot\frac{\theta_0}{2}
\left(t+\sqrt{t^2+1}\right)^p\right].
\label{solitonSolution_th0}
\end{equation}
This is illustrated for $p=1$ in Fig.\rf{solitonFig}.

We focus on the asymptotically planar alignment,
$\theta(r\rightarrow\infty,z)=\theta_0=\pi/2$, corresponding to $t_0=0$. On the physical surface boundary at $z=0$, the tilt angle is then given by a nontrivial function of $r$,
\begin{eqnarray}
\hspace{-0.5cm}
\theta(r,z=0)&=&\theta_0(r)=\theta_s(z_0/r),\\
&=&2\mathrm{arccot}\left[\left(z_0/r+\sqrt{z_0^2/r^2+1}\right)^p\right],
\label{boundary_0}
\end{eqnarray}
that describes the escape into the third dimension [vanishing
$\theta_0(r)$] on the surface $z=0$ inside a disk of radius $z_0$, as illustrated in Fig.\rf{surface_escapeFig}. As anticipated above, $z_0$ allows only a single parameter adjustment of the boundary condition, physically controlled by $V_s$\cite{commentKleman}. An infinitely strong surface anchoring, $V_s\rightarrow\infty$, gives a perfectly planar alignment, $\theta_0(r)=\pi/2$, characterized by $z_0\rightarrow 0$.

We note, however, that a more general boundary condition,
$\theta_0(r)$ can be imposed by generalizing above exact solution to an $r$-dependent $z_0(r)$. Although the resulting
$\theta_s([z+z_0(r)]/r)$ is no longer an exact solution to the E-L equation, for a small $\partial_r z_0$ it is an accurate approximation and can be employed as a good variational ansatz.
\begin{figure}
  \includegraphics[height=2 in]{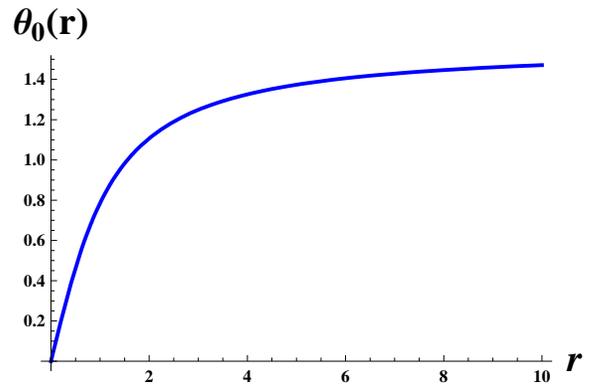}\\       
  \caption{Surface tilt angle $\theta_0(r)$ for a ($p=1$) $2\pi$-vortex, showing surface escape into the third dimension (a meron), confined to a radius $z_0$.}
\label{surface_escapeFig}
\end{figure}

\subsection{Pontryagin index of conical solitons}
\label{Qindex}
The solitonic solutions $\theta_p(\xv)=\theta_s(t)$,
Eq.\rf{solitonSolution_th0} (indexed by $p$), together with
$\phi_p(\xv)=p\varphi$, give the unit director field $\nh_p(\xv)$ according to parametrization Eq.\rf{nhat}. When restricted to a two-dimensional closed surface, e.g., a sphere $S_2^\xv$ in coordinate space $\xv$, $\nh_p(\hat{\xv})$ gives a mapping of this coordinate sphere $S_2^\xv$ into another sphere $S_2^{\nh}$, where $\nh$ ``lives.'' Such mappings fall into topologically distinct classes,
that form the second homotopy group, $H_2(S_s)={\cal{Z}}$,
corresponding to distinct ways of wrapping a coordinate sphere around a target space sphere. The classes are characterized by the Pontryagin topological index
\begin{equation}
Q=\frac{1}{8\pi}
\int d a_k\epsilon_{ijk}\nh\cdot(\partial_i\nh\times\partial_j\nh),
\label{Q}
\end{equation}
where $d a_k$ is the $k$th component of the infinitesimal surface element pointing along the local surface normal.  We have computed $Q$ for our director field solutions $\nh_p(\xv)$, and found that $Q=p$.

\subsection{Weak surface pinning (small $\theta$) analysis}

For weak surface pinning the tilt angle $\theta_0(r)$ is small, corresponding to a large $z_0$, and for a large range $0 < r < z_0$, solution Eq.\rf{solitonSolution_th0} reduces to
\begin{eqnarray}
\theta(t)&=&\frac{\theta_0}{\left(t+\sqrt{t^2+1}\right)^p},\nonumber\\
&=&\theta_0\left(\sqrt{t^2+1}-t\right)^p.
\label{smallTheta}
\end{eqnarray}

We can compare this result with that obtained by studying the
linearized\cite{commentLinear} Euler-Lagrange equation, 
\begin{equation}  
r^2\partial_r^2\theta(r,z)+r\partial_r\theta(r,z)+
r^2\partial_z^2\theta(r,z)-p^2\theta(r,z)=0.
\end{equation}
This differential equation is separable. Letting
$\theta(r,z)=R(r)Z(z)$, it becomes
\begin{equation}
(\frac{R''}{R}+\frac{1}{r}\frac{R'}{R}-\frac{p^2}{r^2})=-\frac{Z''}{Z}
=-k^2, 
\end{equation}
where the sign of the constant $-k^2$ is chosen negative to ensure a well-behaved solution that decays at large $z$.  Keeping only the decaying solution, $Z(z)$ is given by
\begin{equation}
Z(z)=Z_0e^{-kz}.
\end{equation}
Noting that the equation for $R(r)$ is the Bessel equation of order $p$ in variable $k r$, the full solution of the E-L equation for weak pinning is given by
\begin{equation}
\theta(r,z)=\int_0^{\infty} dk a_kJ_p(kr)e^{-kz} \label{smalltheta_general}
\end{equation}
where the coefficients $a_k$ are determined by the boundary condition at $z=0$, namely by $\theta_0(r)=\theta(r,z=0)$. Using the orthogonality of Bessel functions, these are given by
\begin{equation}
a_k=k\int_0^{\infty} dr r\theta(r,0)J_p(kr).
\label{ak}
\end{equation}
We note that, in contrast to the full solitonic solution,
Eq.\rf{solitonSolution} [where we were only able to impose a boundary condition with a specific $r$ dependence, $\theta_0(r)=\theta_s(z_0/r)$, displayed in Fig.\rf{surface_escapeFig}], here we can obtain a solution $\theta(r,z)$ for an arbitrary $r$-dependent boundary condition $\theta(r,z=0)=\theta_0(r)$.

To compare to the full solution, we choose a constant boundary
condition $\theta_0$, for which
\begin{eqnarray}
a_k&=&k\int_0^{\infty} dr r\theta_0J_p(kr),\\
&=&\theta_0\, p/k, 
\end{eqnarray}
where a convergence factor $e^{-0^+kr}$ had to be introduced to make the integral into $\int_0^\infty dx x J_p(x) e^{-0^+x}$, which is well defined and equal to $p$. Using these expansion coefficients $a_k$ inside Eq.\rf{smalltheta_general} we find
\begin{eqnarray}
\theta(r,z)
&=&\theta_0 p\int_0^{\infty}\frac{dk}{k}J_p(kr)e^{-kz},\\
&=&\theta_0\left(\sqrt{\frac{z^2}{r^2}+1}-\frac{z}{r}\right)^p,
\end{eqnarray}
in complete agreement with the small $\theta_0$ limit [Eq.\rf{smallTheta}] of the full solitonic solution.

\section{Energetics}

\subsection{Conical soliton energy}

The bulk elastic energy corresponding to the surface vortex solution, $\theta_s(t)$, found above is straightforwardly computed by plugging into the elastic Hamiltonian and evaluating the spatial integrals. We thereby obtain
\begin{eqnarray}
E^{(p)}_s(\theta_0,z_0)&\equiv&H_{el}[\theta_s(t)],\\
&=&\half K\int d^2\rv dz \bigg[(\nabla\theta_s)^2 
+ \frac{p^2}{r^2}\sin^2\theta_s\bigg],\;\;\;\;\;\;\;\\
&=&2\pi p^2 K\int_0^{L_r} dr \int_{z_0/r}^\infty dt \sin^2\theta_s,
\label{Est0}
\end{eqnarray}
where we took advantage of the energy integral of ``motion,''
Eq.\rf{conserveE2} to eliminate $(\nabla\theta_s)^2$, and $L_r$ is the extent of the system in the radial direction. Using the explicit solution for $\theta_s(t)$ and defining
\begin{eqnarray}
x(t)&=&t+\sqrt{t^2+1},\nonumber\\
x_0&\equiv& x(t_0)=\left(\tan(\theta_0/2)\right)^{1/p},
\label{xt}
\end{eqnarray}
we obtain 
\begin{eqnarray}
E^{(p)}_s(\theta_0,z_0)&=&
2\pi K \int_0^{L_r} dr 
\varepsilon(x_0,x(z_0/r),p),
\label{Esx_p}
\end{eqnarray}
where
\begin{equation}
\varepsilon(x_0,x(z_0/r),p)
=2p^2\int_{x(z_0/r)}^\infty
dx\frac{x^2+1}{x^2\big((x/x_0)^p+(x_0/x)^p\big)^2}.
\label{eps_p}
\end{equation}

\subsubsection{p=1 vortex energy}
Specializing to the case of $p=1$ surface vortex, above energy
is simplified and can be calculated analytically 
\begin{eqnarray}
E^{(1)}_s(\theta_0,z_0)&=&
4\pi K \int_0^{L_r} dr 
\int_{x(z_0/r)}^\infty dx x_0^2\frac{x^2+1}{(x^2+x_0^2)^2},\nonumber\\
&\equiv&2\pi K \int_0^{L_r} dr \varepsilon(x_0,x(z_0/r),1),\\
&\equiv&2\pi K z_0\int_{z_0/L_r}^\infty dt
t^{-2}\varepsilon(x_0,x(t)),
\label{Esx0c}
\end{eqnarray}
where
\begin{eqnarray}
\varepsilon(x_0,x,1)&=&
\frac{\pi}{2}x_0 +  x\frac{x_0^2-1}{x_0^2+x^2} 
- x_0\ \mathrm{arccot}\frac{x_0}{x} + 
\frac{1}{x_0}\mathrm{arctan}\frac{x_0}{x}.\nonumber\\
\end{eqnarray}
For a vanishing $z_0$, $x=x(z_0/r)=1$, and we find
\begin{eqnarray}
E^{(1)}_s(\theta_0,0)&=&2\pi K L_r\varepsilon(x_0,1,1),
\label{Esx0}
\end{eqnarray}
where
\begin{eqnarray}
\varepsilon(x_0,1,1)&=&
\frac{\pi}{2}x_0 +  \frac{x_0^2-1}{x_0^2+1^2} 
- x_0\ \mathrm{arccot}x_0 + 
\frac{1}{x_0}\mathrm{arctan}x_0\,\nonumber\\
&&\\
&\approx&
\begin{cases} \frac{8}{3} x_0^2,
& \text{for $x_0\ll 1$}\cr
\frac{\pi}{2}+2(x_0-1),
& \text{for $x_0\rightarrow 1^{-}$}.\cr
\end{cases}
\label{epsilonx0}
\end{eqnarray}
Using the relation $\theta_0(x_0)$ [Eq.\rf{xt}] to express the soliton energy in terms of the surface tilt angle $\theta_0$, we obtain the energy $E^{(1)}_s(\theta_0,0)$ of a surface $2\pi$-vortex plotted in Fig.\rf{EsFig}.
\begin{figure}
  \includegraphics[height=2.2in]{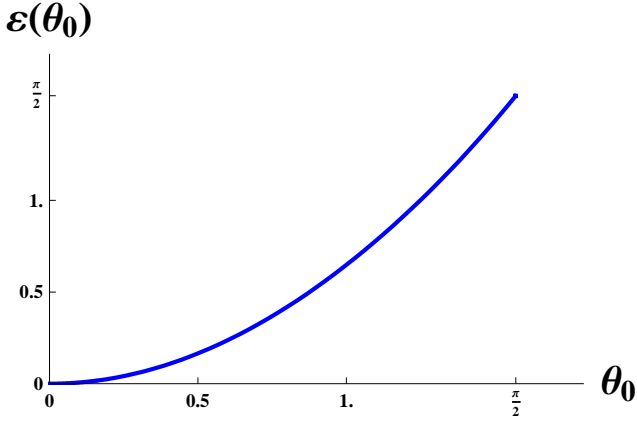}\\       
  \caption{Energy $\varepsilon(\theta_0,1,1)$ (in units of $2\pi K L_r$) of the soliton texture due to a $2\pi$ surface vortex as a function of the surface tilt angle $\theta_0$, together with its parabolic approximation, $\varepsilon\approx a\theta_0^2$, with $a$ fitted to be $0.64$ (appearing indistinguishable).}
\label{EsFig}
\end{figure}

For asymptotic planar alignment $\theta_0(r\rightarrow\infty)=\pi/2$ ($t_0=0$ and $x_0=1$)
\begin{eqnarray}
\varepsilon(1,x)&=&
\frac{\pi}{2} - \mathrm{arctan}x + 
\mathrm{arccot}x.
\end{eqnarray}
Substituting this into Eq.\rf{Esx0c} we find
\begin{eqnarray}
E^{(1)}_s(\pi/2,z_0)&=&2\pi K L_rg(z_0/L_r),
\end{eqnarray}
where the scaling function $g(\zh_0)$ is given by 
\begin{eqnarray}
g(\zh_0)&=&\frac{\pi}{2}+\mathrm{arccot}\left(\zh_0+\sqrt{\zh_0^2+1}\right)\nonumber\\
&&-\mathrm{arctan}\left(\zh_0+\sqrt{\zh_0^2+1}\right)+\frac{1}{2}\zh_0\ln\left(\frac{\zh_0^2}{1+\zh_0^2}\right).
\end{eqnarray}

\subsubsection{p vortex energy}

\begin{figure}
  \includegraphics[height=2.2 in]{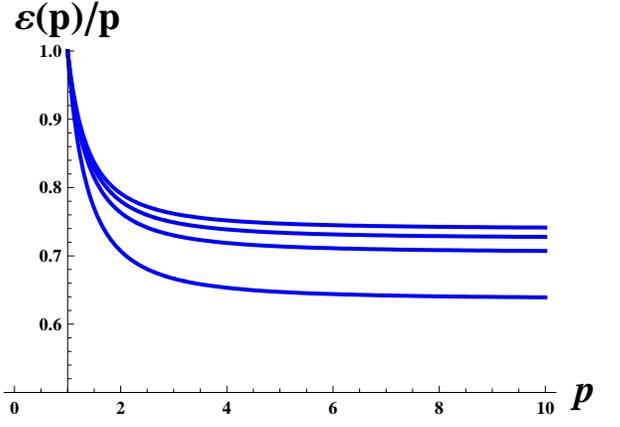}\\       
  \caption{Energy $\varepsilon(\theta_0,1,p)/p$ [in units of 
$2\pi K L_r \varepsilon(\theta_0,1,1)$] of the soliton texture due to a $2\pi p$ surface vortex as a function of its topological charge $p$, displayed for surface tilt angles $\theta_0=\pi/2, \pi/3, \pi/4, \pi/6$ (top to bottom).}
\label{fig:eps_p}
\end{figure}

For $p>1$ charge vortex, energy $E^{(p)}_s$ can only be evaluated numerically. Focusing on $z_0=0$ for simplicity,
\begin{eqnarray}
E^{(p)}_s(\theta_0,0)&=&
2\pi K L_r\varepsilon(x_0,1,p),
\label{Es1_p}
\end{eqnarray}
where we evaluated $\varepsilon(x_0,1,p)$ numerically and displayed $\varepsilon(x_0,1,p)/p$ as a function of charge $p$ for various values of $x_0$ in Fig.~\ref{fig:eps_p}. As can be seen from this figure, despite the fact that the naive $p$ dependence of $\varepsilon(x_0,1,p)$ in Eq.~\ref{eps_p} is the standard $p^2$ found in a 2D vortex, the $p$ dependence coming from the integral reduces it to an asymptotically {\em linear} one at large $p$, 
\begin{equation}
\varepsilon(x_0,1,p\gg 1)\approx (1-\cos\theta_0)p.
\end{equation}
This finding has important qualitative implication that one
winding-$p$ surface vortex (one $p$-boojum) has a lower energy than $p$ winding-$1$ surface vortices ($p$ $1$-boojums). This contrasts strongly with the standard 2D vortex case where a $p$-vortex always has a higher energy than $p$ unit-vortices and thus always fissions into them.

\subsection{Competing states}

We can compare the energy of the solitonic state,
$\theta_s[(z+z_0)/r]$, discussed above with competing states
illustrated in Fig.\rf{line_wallFig}. To this end, we estimate
energetics by simple scaling analysis for a system of size
$L_{r}\times L_z$, focusing on the strong planar alignment,
$\theta_0=\pi/2$.

\begin{figure}
\hspace{-0.3cm}\includegraphics[width=3.5 in]{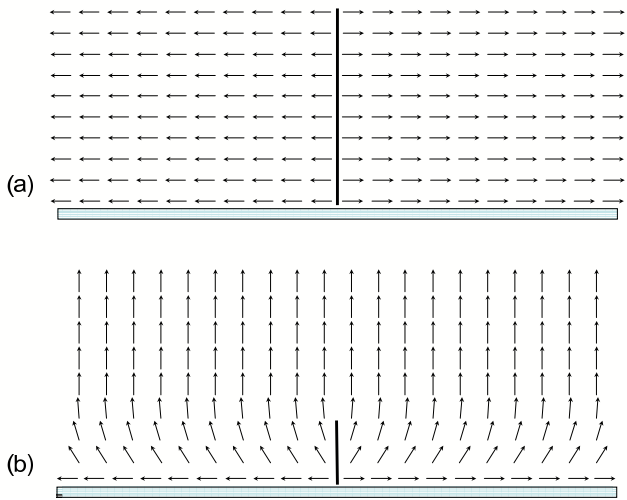}\\       
\caption{Competing states for a surface $2\pi$-vortex boundary
  condition, with (a) vortex line and (b) domain-wall texture
  extensions into the bulk.}
\label{line_wallFig}
\end{figure}

\subsubsection{Vortex line}
One competing configuration is the ``vortex line'' state that extends the surface $2\pi$-vortex into a straight vortex line with $\theta(z)=\theta_0=\pi/2$, independent of $z$.  The energy of such a state is clearly given by
\begin{equation}
E_{vortex line}=
\frac{K}{2}\int \frac{1}{r^2}2\pi r dr dz \propto K L_z\ln{\frac{L_r}{a}},
\end{equation}
where $a$ is the core radius of the vortex line, set by the coherence
length. 

\subsubsection{Domain wall}
Another possible texture is that of a 2D domain wall of thickness $a$, where $\theta(z)$ exhibits a uniform (i.e., $r$ independent) escape into the third dimension, changing from $\theta_0=\pi/2$ to $\theta=0$ over a microscopic distance $a$. The corresponding energy is given by
\begin{equation}
E_{domain wall}=\frac{K}{2}\int_0^{L_r} 2\pi rdr\int_0^a dz
\left(\frac{\pi/2}{a}\right)^2\propto KL_r^2/a, 
\end{equation}
scaling with the area of the cell.

\subsubsection{Conical soliton surface vortex}
The energy of a conical soliton surface vortex can be similarly estimated. We first note that by virtue of the E-L equation, all three ($z$ derivatives, $r$ derivatives, and $\sin^2\theta$) contributions are comparable, and therefore we can focus on one of them. Estimating the elastic energy based on the $z$ derivatives, we observe that the strain is confined to a soliton width along $z$ that at radius $r$ is
given by $\xi_z\approx r$. Thus the estimate is quite similar to the previous case of the domain wall but with strain spread out over region between the cones $z=r$ and $z=0$ rather than confined to a slab $0<z<a$. This leads to an estimate
\begin{equation}
E_s\approx K\int_0^{L_r}2\pi r dr 
\int_0^{r} dz \left(\frac{\pi/2}{r}\right)^2\propto KL_r,
\label{Es_est}
\end{equation}
which agrees qualitatively with our exact computation Eq.\rf{Esx0}.

Since the conical soliton solution scales only {\em linearly} in $L_r$, we conclude that the domain-wall solution (scaling as $L_r^2$), is not competitive with the other two solutions. On the other hand, the relative competition between the vortex line and conical soliton solution depends on the relative ratio of $L_z$ and $L_r$. 

For $L_z > L_r$ clearly vortex line is energetically more costly and conical soliton texture is the preferred state. On the other hand for a thin cell with width $L_z=w < L_r$ a more detailed analysis is required. The vortex line energy is still clearly given by $E_{vortex line} = K w\ln\frac{L_r}{a}$.

To compute a conical soliton energy in a cell of a finite width $w$ requires an extension of the solution to a finite geometry. For a finite width cell with {\em free} and planar boundary conditions on the top and bottom substrates, respectively, our exact solution, $\theta_s(t)$ is a good description. Its energy can be simply estimated. Examining Fig.\rf{SVsolution2DFig}, it is clear that for $w< L_r$, there are two additive energy contributions of this
texture. For the region $0<r< w$, the contribution is identical to that made in Eq.\rf{Es_est}. On the other hand, for region $r > w$, the strain field is that of a $2\pi$-vortex line with length $w$ and core radius $w$. Putting these two contributions together, we find
\begin{eqnarray}
E_{conic-soliton}&\approx& K w + K w\ln\frac{L_r}{w},\\
&\approx& K w (1 + \ln\frac{L_r}{a} - \ln\frac{w}{a}) <
E_{vortex-line},
\nonumber\\
&&
\label{Econic_wins}
\end{eqnarray}
for cell thickness $w \gg a$.

\begin{figure}[bth]
\centering
\setlength{\unitlength}{1mm}
\begin{picture}(100,70)(0,0)
\put(-4,-9){\begin{picture}(100,85)(0,0)
\includegraphics{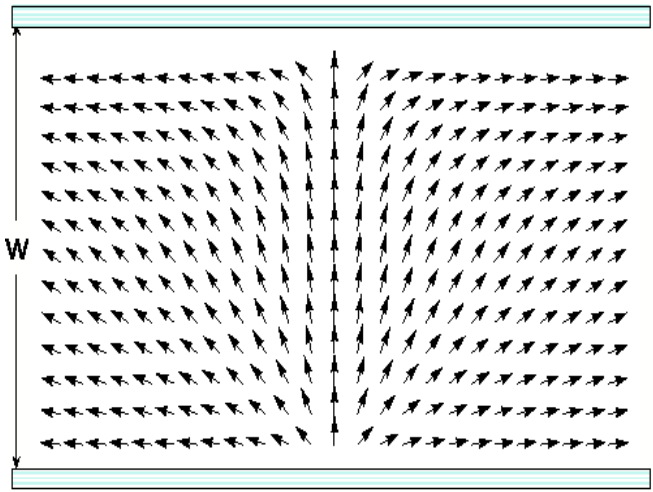}
\end{picture}}
\end{picture}
\caption{A solitonic texture describing escape into a third dimension of two $2\pi$-vortices confined to top and bottom substrates of a finite width cell.}
\label{w_solitonFig}
\end{figure}


Unfortunately, we have been unable to find an exact solution for the experimentally more relevant case of {\em non-free} (e.g., {\em symmetric planar}) boundary conditions on both substrates. The difficulty has to do with the failure of a periodic soliton solution [obtained by picking the integration constant in Eq.\rf{conserveE2} to be $E<1$ and matching its period to the width of the cell; see the
Appendix] to enforce fixed $z$ (as opposed to fixed $t$) boundary conditions.

However, a good approximate symmetric solution, illustrated in
Fig.\rf{w_solitonFig}, is given by
\begin{equation}
\theta_s^{w}(r,z)=\theta_s((w/2-|z|)/r).
\label{w_solitonSolution}
\end{equation}
Describing a cell with two planar aligning substrates at $z=\pm w/2$, its only shortcoming is a small slope discontinuity in the $z$ derivative at $z=0$ (the center plane of the cell).


\acknowledgments

We thank I. Smalyukh, V. Gurarie and S. Choi for discussions and comments on the paper. The authors acknowledge financial support by the National Science Foundation through Grants No.\ DMR-0321848 and No.\ MRSEC DMR-0213918.

\appendix
\section*{Extension to a periodic and finite width solution}
\label{periodic}

The analogy of the E-L equation [Eq.\rf{conserveE2}] with a
fictitious particle dynamics allows an extension of the single soliton solution to a periodic soliton array.  The latter is obtained by choosing the integration constant $E < 1$, corresponding to the particle starting with a vanishing velocity and below the potential maximum. The subsequent ``evolution'' of $\theta_s(t)$ is clearly periodic in $t$, confined to the range $0 \leq \theta_s(t) \leq \theta_m$, with $\theta_m=\frac{1}{2}\mathrm{arccot}(E)$. 

Going back to Eq.\rf{integralSoln} we observe that the $\theta'$ integral can be related to the Legendre form of the elliptic integral of the first kind,
\begin{equation}
F(\phi,k)=\int_0^\phi\frac{d\phi'}{\sqrt{1-k^2\sin^2\phi'}}.
\end{equation}
Thus our solution can be expressed in terms of the Jacobi elliptic function $\mathrm{sn}(\phi,k)$ defined by
\begin{eqnarray}
\sn[F(\phi,k),k]&=&\sin\phi.
\end{eqnarray}
$\sn(t,k)$ is an odd periodic function resembling a
smoothed out square wave. For $k>1$ it interpolates between a single soliton for $k=1^+$ (half a period of a square wave) and $k^{-1}\sin k t$ for $k\gg1$. For $k > 1$ the period of $\sn(t,k)$ is given by $2F[\sin^{-1}(1/k),k]$.

To establish a direct relation we change variables $\phi' = \theta'-\pi/2$, finding
\begin{eqnarray}
F(\phi,k)&=&\sn^{-1}(\sin\phi,k),\\
&=&\int_{\pi/2}^{\phi+\pi/2}\frac{d\theta'}{\sqrt{1-k^2\cos^2{\phi'}}},\\
&=&\frac{\sqrt{2}}{k}
\int_{\pi/2}^{\phi+\pi/2}\frac{d\theta'}{\sqrt{E_k- \cos{2\theta'}}},
\label{sn_define}
\end{eqnarray}
where $E_k = (2-k^2)/k^2$. In this notation, Eq.\rf{integralSoln} becomes
\begin{eqnarray}
\int_{t_0}^tdt'\frac{1}{\sqrt{t'^2+1}} &=& 
-\sqrt{2}\int_{\pi/2}^\theta\frac{d\theta'}{\sqrt{E-\cos2\theta'}},\\
\ln\left[\frac{t+\sqrt{t^2+1}}{t_0+\sqrt{t_0^2+1}}\right]
&=&-k_E\sn^{-1}[\sin(\theta-\pi/2),k_E],\hspace{1cm}\\
&=&k_E\sn^{-1}[\cos\theta,k_E],\\
&&\nonumber
\label{sn_soln}
\end{eqnarray}
where $k_E=\sqrt{2/(1+E)}$, and we used the fact that $\sn[\phi,k]$ is an odd function of $\phi$.  Thus the periodic conical soliton solution is given by
\begin{eqnarray}
\hspace{-0.5cm}
\theta_s(t,k)=\arccos\left\{\sn\left[\frac{1}{k}
\ln\bigg(\frac{t+\sqrt{t^2+1}}{t_0+\sqrt{t_0^2+1}}\bigg),k\right]\right\},
\label{sn_soln2}
\end{eqnarray}
with $k=1^+$ giving our earlier single soliton solution [Eq.\rf{solitonSolution}].

One might hope to use this solution to model a finite thickness, $w$, {\em symmetric} liquid-crystal cell with two boundaries inducing a symmetric (about $z=w/2$) director rotation from $\theta=\pi/2$ to $0$ and back to $\pi/2$. Naively, this maybe done by choosing the value of $k$ such that the period matches the cell thickness, $w$. Although
this is possible for standard 1D solitonic problems, because here the solution is periodic in $t=(z+z_0)/r$ (stemming from the fact that we are dealing with a 2D problem) and {\em not} in $z$, solution \rf{sn_soln2} cannot be used to model a cell with {\em symmetric} boundaries at fixed $z=0$ and $z=w$.  A more general class of solutions is necessary but is currently unavailable.


\begin{thebibliography}{99}
\bibitem{commentKleman} In Ref.\onlinecite{Kleman} it was demonstrated
  that for a special (but realistic) form of the surface potential,
  $V(\theta)=V_s\cos^2\theta$, the boundary condition for the $p=1$ surface
  defect can be satisfied exactly.
\end{thebibliography}
\end{document}